\documentclass[aps,prb,reprint,amsmath,amssymb,superscriptaddress]{revtex4-2}
\usepackage{graphicx,xcolor}

\begin{document}

\title{Why polar excitons stay sharp: parity protection of the center-of-mass recoil channel in exciton-phonon scattering}

\author{Michael O. Atambo}
\email{michael.atambo@tukenya.ac.ke}
\affiliation{Department of Physics, Earth and Environmental Science, Technical University of Kenya, Nairobi, Kenya}

\date{\today}

\begin{abstract}
In polar semiconductors the Fr\"ohlich interaction is the dominant
electron--phonon coupling, yet excitonic resonances in materials such as
halide perovskites remain anomalously sharp. We show that standard
frozen-center-of-mass treatments of the exciton--phonon problem miss the
decisive kinematic degree of freedom: restoring the exact center-of-mass
(COM) recoil reveals a universally open, parameter-free $1s\to1s$
absorption channel at recoil momentum $q_*=\sqrt{2M_{\rm ex}\hbar\omega_{\rm
LO}}/\hbar$, whose rate scales as $N_{\rm LO}(T)$. We prove that this
recoil channel is controlled by destructive electron--hole interference:
the recoil linewidth vanishes with the mass asymmetry as
$\gamma_{\rm LO}^{\rm recoil}\propto\mathcal{F}_{1s,1s}(q_*)^2$, and the
elastic dressing obeys the exact suppression law
$S_X/S_{\rm ind}=\eta^2(6-\eta^2)/5$ within the hydrogenic Fr\"ohlich
model. The theory establishes a hierarchy of scattering regimes. In
mass-asymmetric materials (GaAs, $\eta=-0.74$) the recoil channel is
active ($\gamma_{\rm LO}^{\rm recoil}=2.2$~meV); in mass-symmetric
materials (FAPbI$_3$, $\eta=0$; MAPbI$_3$, $\eta=-0.11$) it is killed by
interference (0.00 and 0.12 meV), showing that the observed 27--40 meV
perovskite linewidths cannot be accounted for by COM recoil and therefore
require internal-state-changing and other inelastic channels, of which the
constructive, $\eta$-robust $1s\to np$ resonance is the leading candidate
within the present model. The Fr\"ohlich constant $\alpha$ alone is
therefore insufficient as a figure of merit: after projection onto the
correlated exciton, the controlling parameters are $\eta$, $q_*a_X$, and
the Rydberg detuning.
\end{abstract}

\maketitle

\section{Introduction}

The interaction of excitons with polar optical phonons governs the
coherence of optical excitations in semiconductors, from conventional
III--V compounds to halide perovskites and monolayer
chalcogenides~\cite{Frohlich1937,Toyozawa1958,RudinReineckeSegall1990}.
The standard description treats the electron and hole as independent
Fr\"ohlich polarons~\cite{Frohlich1954}, an approach that predicts strong
lattice dressing whenever the Fr\"ohlich constant $\alpha$ is large. Yet
materials with $\alpha\sim2$--3, such as lead-halide perovskites, exhibit
sharp zero-phonon lines and modest homogeneous
broadening~\cite{Wright2016,Fu2018}, a discrepancy that has motivated
extensive phenomenological fitting of temperature-dependent
linewidths~\cite{RudinReineckeSegall1990,Fu2018}.

The independent-polaron picture neglects that a phonon couples to the
\emph{combined} charge density of a correlated electron--hole pair.
Many-body perturbation theory (MBPT) in the $GW$--Bethe--Salpeter (BSE)
framework~\cite{BetheSalpeter,Hedin1965,HybertsenLouie1986,Strinati1980,Strinati1982,
OnidaReiningRubio2002,AryasetiawanGunnarsson1998} provides the natural
language for such correlated excitations, and recent work has begun to
construct exciton--phonon self-energies beyond independent
quasiparticles. In Ref.~\cite{Atambo2026a} we derived the exact
multipole structure of the excitonic Fr\"ohlich vertex and established a
parity-controlled interference selection rule; in Ref.~\cite{Atambo2026b} we quantified, against exact diagonalization, when
the underlying $GW$ description itself may be trusted.

Here we derive the observable spectral consequences of that selection rule
using only analytically tractable models, and establish three results.
(i) The internal-state-preserving vertex is infrared-safe and the
Huang--Rhys factor obeys an exact suppression law, Eq.~(\ref{eq:R}).
(ii) The inelastic vertex is finite, dipolar, and robust against mass
asymmetry. (iii) The homogeneous linewidth is kinematically gated: the
recoil coefficient $\gamma_{\rm LO}^{\rm recoil}$ is a parameter-free
number fixed by $(\eta,\ q_*a_X)$, while the internal-state-changing
channels are gated by the detuning
$|\Delta E_{2p}-\hbar\omega_{\rm LO}|$. Together these results explain why
polar excitons stay sharp and identify the design parameters that control
exciton--phonon scattering.

\section{Theory}

\subsection{Exciton-projected Fr\"ohlich vertex}

We describe the relative motion of the exciton by hydrogenic envelope
functions $\psi_{nlm}(\mathbf r)$ with Bohr radius $a_X$, and couple it to a
dispersionless LO mode through the Fr\"ohlich vertex
$M(q)=i[2\pi e^2\hbar\omega_{\rm LO}/(Vq^2)]^{1/2}
(\varepsilon_\infty^{-1}-\varepsilon_s^{-1})^{1/2}$~\cite{Frohlich1937}.
Projecting onto the relative coordinate, the electron and hole sample
momenta $f_e\mathbf q$ and $-f_h\mathbf q$, where
$f_{e,h}=m_{h,e}/(m_e+m_h)$ are the mass fractions. The excitonic vertex is
\begin{equation}
G_{SS'}(q)=M(q)\,\langle S'|e^{if_e\mathbf q\cdot\mathbf r}
-e^{-if_h\mathbf q\cdot\mathbf r}|S\rangle\equiv M(q)\,
\mathcal F_{SS'}(q).
\label{eq:vertex}
\end{equation}
All physics below follows from the excitonic form factors
$\mathcal F_{SS'}(q)$, both their small-$q$ multipole structure and their
finite-$q$ values at the recoil momentum $q_*$.

\subsection{Internal-state-preserving channel: exact parity protection}

We first consider scattering processes that preserve the internal excitonic quantum numbers ($1s\to1s$) but involve a change in the center-of-mass (COM) momentum. The excitonic form factor is
\begin{equation}
\mathcal F_{1s,1s}(q) = \langle 1s | e^{i f_e \mathbf q\cdot\mathbf r} - e^{-i f_h \mathbf q\cdot\mathbf r} | 1s \rangle.
\label{eq:F1s1s}
\end{equation}
For a mass-symmetric exciton
($m_e=m_h\Rightarrow\eta=0$)\footnote{We define
$\eta=(m_e-m_h)/(m_e+m_h)$, opposite in sign to the convention of
Ref.~\cite{Atambo2026a}; all rates depend on $\eta^2$ and
$|F_{1s,1s}(q_*)|$ and are unaffected by this choice.}, the mass
fractions are exactly $f_e=f_h=1/2$. Eq.~(\ref{eq:F1s1s}) then reduces to
\begin{equation}
\mathcal F_{1s,1s}(q, \eta=0) = 2i \langle 1s | \sin(\mathbf q\cdot\mathbf r / 2) | 1s \rangle.
\label{eq:exact_zero}
\end{equation}
Because the hydrogenic $1s$ envelope is purely real and spherically symmetric (even parity), and $\sin(\mathbf q\cdot\mathbf r / 2)$ is an odd function of $\mathbf r$, the integrand is strictly odd. Consequently, $\mathcal F_{1s,1s}(q, \eta=0) = 0$ \emph{identically for all magnitudes of $q$}. This is a much stronger result than the small-$q$ expansion: the internal-state-preserving coupling to the macroscopic polar field is exactly forbidden by parity for any mass-symmetric exciton, regardless of the recoil momentum.
For later use we record the closed-form hydrogenic result
$\langle e^{i\mathbf k\cdot\mathbf r}\rangle_{1s}=(1+k^2a_X^2/4)^{-2}$, whence
\begin{equation}
\mathcal F_{1s,1s}(q)=(1+f_e^2x^2/4)^{-2}-(1+f_h^2x^2/4)^{-2},\quad x\equiv qa_X,
\label{eq:closedform}
\end{equation}
which vanishes identically for $f_e=f_h$ (all $q$). The exact symbolic
multipole residual of Eq.~(\ref{eq:closedform}) is
\begin{equation}
\mathcal F_{1s,1s}(q)=\frac{\eta}{2}(qa_X)^2
-\frac{3\eta(1+\eta^2)}{32}(qa_X)^4+\mathcal O(q^6),
\label{eq:multipole}
\end{equation}
so that the monopole and dipole vanish identically and the leading
quadrupole term removes the $1/q$ Fr\"ohlich divergence.

For mass-asymmetric excitons ($\eta \neq 0$), the cancellation is
incomplete. The Huang--Rhys~\cite{HuangRhys1950} factor
$S_X=\sum_q|G_{1s,1s}(q)|^2/(\hbar\omega_{\rm LO})^2$ is infrared-finite
and, evaluating the integral in closed form (Sec.~III), obeys the exact
suppression law
\begin{equation}
S_X=R(\eta)\,S_{\rm ind},\qquad
R(\eta)=\frac{\eta^2(6-\eta^2)}{5},
\label{eq:R}
\end{equation}
where $S_{\rm ind}=S_e+S_h$ is the independent-polaron
result~\cite{Frohlich1954}. $R(\eta)$ is exact for all $\eta$: it reduces
to $C\eta^2$ with $C=6/5$ at small $\eta$ and saturates at $R(1)=1$, the
limit in which a heavy point-like carrier cannot be canceled.

\textit{Relation to other work.} This work extends the leading-order parity selection rule of Ref.~\cite{Atambo2026a} to an
exact all-$q$ identity, Eq.~(\ref{eq:closedform}); Ref.~\cite{Atambo2026a} focuses on the inelastic multipole expansion, the dressing phase diagram, and the 2D extensions.

\subsection{On-shell linewidths: COM recoil and inelastic resonance}

A common approximation freezes the exciton COM at $\mathbf K=0$, artificially closing the $1s\to1s$ absorption channel for a dispersionless LO mode. Restoring the COM dispersion $E_X(\mathbf K) = E_{1s} + \hbar^2 K^2 / 2M_{\rm ex}$ reveals a universally open, internal-state-preserving recoil channel (inelastic in total energy, although the internal state is preserved). An exciton at $\mathbf K=0$ may absorb an LO phonon and transition to $1s$ at finite momentum $\mathbf q$. Energy conservation dictates
\begin{equation}
\hbar\omega_{\rm LO} = \frac{\hbar^2 q^2}{2M_{\rm ex}} \quad\Rightarrow\quad q_* = \frac{\sqrt{2M_{\rm ex}\hbar\omega_{\rm LO}}}{\hbar}.
\label{eq:recoil}
\end{equation}
The absorption rate is derived from Fermi's Golden Rule:
\begin{equation}
\Gamma_{\rm abs} = \frac{2\pi}{\hbar} \frac{V}{(2\pi)^3} \int d^3q \, |G_{1s,1s}(\mathbf q)|^2 N_{\rm LO}(T) \delta\left(\frac{\hbar^2 q^2}{2M_{\rm ex}} - \hbar\omega_{\rm LO}\right).
\end{equation}
Converting the delta function using $\delta(f(q)) = \delta(q-q^*) / |f'(q^*)|$ yields the Jacobian $|f'(q^*)|^{-1} = M_{\rm ex} / (\hbar^2 q^*) = 1 / (\hbar v_*(q^*))$, where $v_*$ is the exciton group velocity at $q_*$. Writing $|M(q)|^2=|\tilde M(q)|^2/V$ with
$|\tilde M(q)|^2=2\pi e^2\hbar\omega_{\rm LO}(4\pi\varepsilon_0q^2)^{-1}
(\varepsilon_\infty^{-1}-\varepsilon_s^{-1})$, the volume $V$ from
$\sum_{\bf q}$ cancels the $1/V$ exactly, so the rate is intensive. Performing the angular integration and evaluating the matrix elements on-shell gives the parameter-free recoil linewidth $\Gamma_{\rm recoil}(T) = \gamma_{\rm LO}^{\rm recoil} N_{\rm LO}(T)$, with
\begin{widetext}
\begin{equation}
\gamma_{\rm LO}^{\rm recoil} = \frac{2\pi}{\hbar} \frac{q_*^2}{2\pi^2} \frac{|\tilde M(q_*)|^2 |\mathcal F_{1s,1s}(q_*)|^2}{\hbar v_*(q_*)}.
\label{eq:recoilrate}
\end{equation}
\end{widetext}
\textit{Invariant weight and status of the lattice-gauge split.} The
inclusive coupling weight
$W(q)=\langle 1s|\hat O^\dagger\hat O|1s\rangle=2-2(1+q^2a_X^2/4)^{-2}$,
with $\hat O=e^{if_e\mathbf q\cdot\mathbf r}-e^{-if_h\mathbf q\cdot\mathbf r}$,
is a ground-state expectation value of the physical density operator and is
therefore representation-independent; by the completeness relation it equals
the sum of $|\mathcal F_{1s,S'}(q)|^2$ over all final excitonic states, of
which the intraband term $|\mathcal F_{1s,1s}(q)|^2$ used throughout is one
share. Disagreements between bookkeepings are therefore disagreements about
the intraband share of $W(q)$, never about $W(q)$ itself. For parabolic
bands the center of mass separates exactly (Galilean invariance), the
relative-coordinate vertex of Eq.~(\ref{eq:vertex}) is the physical
operator, and the continuum intraband coefficient $\tfrac12\eta$ used
throughout is exact. The lattice COM-gauge coefficient
$\tfrac32\eta$~\cite{Atambo2026c} is obtained from a
signed-minimum-image vertex which, for $q=2\pi n/L$, $n\neq0$, is not the
periodic lattice density operator; its replacement by the periodic vertex,
and the resulting intraband share, are under audit in the revision of
Ref.~\cite{Atambo2026c} (cluster ledger: OPEN). The rates in
Table~\ref{tab:mat} are continuum-partition values, exact for parabolic
bands; the inclusive weight is unaffected by this open item.
Spontaneous emission is strictly forbidden as $E_X(\mathbf q) \ge E_X(0)$, ensuring $\Gamma \to 0$ as $T\to 0$.

When $\eta \to 0$, Eq.~(\ref{eq:exact_zero}) forces $\gamma_{\rm LO}^{\rm recoil} = 0$. The linewidth is then dictated by the inelastic, internal-state-changing channel ($1s \to np$). The dipole form factor $\mathcal F_{1s,np}(q\to 0) = i q d_{np}$ is constructive and $\eta$-robust.
The internal-state-changing channel is the constructive companion of
Eq.~(\ref{eq:closedform}): expanding the vertex to first order,
$e^{if_e\mathbf q\cdot\mathbf r}-e^{-if_h\mathbf q\cdot\mathbf r}
\approx i(f_e+f_h)\mathbf q\cdot\mathbf r=i\mathbf q\cdot\mathbf r$ since
$f_e+f_h\equiv1$, hence
\begin{equation}
\mathcal F_{1s,np}(q\to0)=i\,\mathbf q\cdot\langle np|\mathbf r|1s\rangle
\equiv i q d_{np},
\label{eq:dipole}
\end{equation}
independent of $\eta$ to leading order.
The corresponding resonant linewidth is
\begin{equation}
\gamma_{\rm LO}^{\rm inel} = \frac{2\pi}{\hbar} \sum_{n>1} \int \frac{d^3q}{(2\pi)^3} |G_{1s,np}(\mathbf q)|^2 \delta_\gamma(\Delta E_{np} + E_{\rm recoil} - \hbar\omega_{\rm LO}),
\label{eq:inelastic}
\end{equation}
where $\delta_\gamma$ accounts for the finite LO phonon lifetime. This channel is gated by the Rydberg detuning $\Delta E_{np} - \hbar\omega_{\rm LO}$. Finally, if $\hbar\omega_{\rm LO} > E_B$ (as in GaAs), the exciton may be ionized ($1s \to \text{continuum}$), adding a rate $\gamma_{\rm ion}$.

\section{Universal suppression of the phonon cloud}

\begin{figure}
\includegraphics[width=\columnwidth]{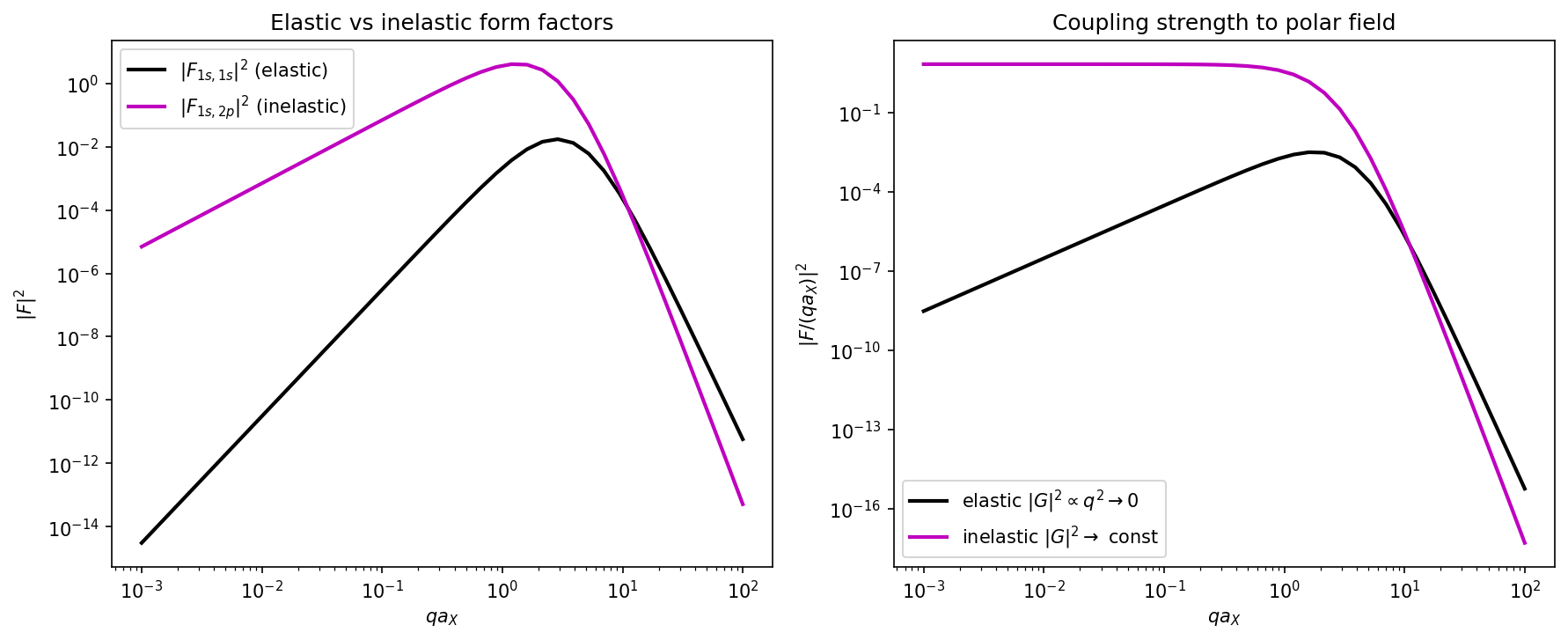}
\caption{(Color online) Internal-state-preserving (black) and
internal-state-changing (magenta) excitonic form factors.
Left: $|\mathcal F|^2$; right: coupling strength
$|\mathcal F/(qa_X)|^2$ to the polar field. The
internal-state-preserving vertex vanishes as $q^2$ (parity
protection); the internal-state-changing vertex tends to a finite
dipolar constant.}
\label{fig:formfactors}
\end{figure}

\begin{figure}
\includegraphics[width=\columnwidth]{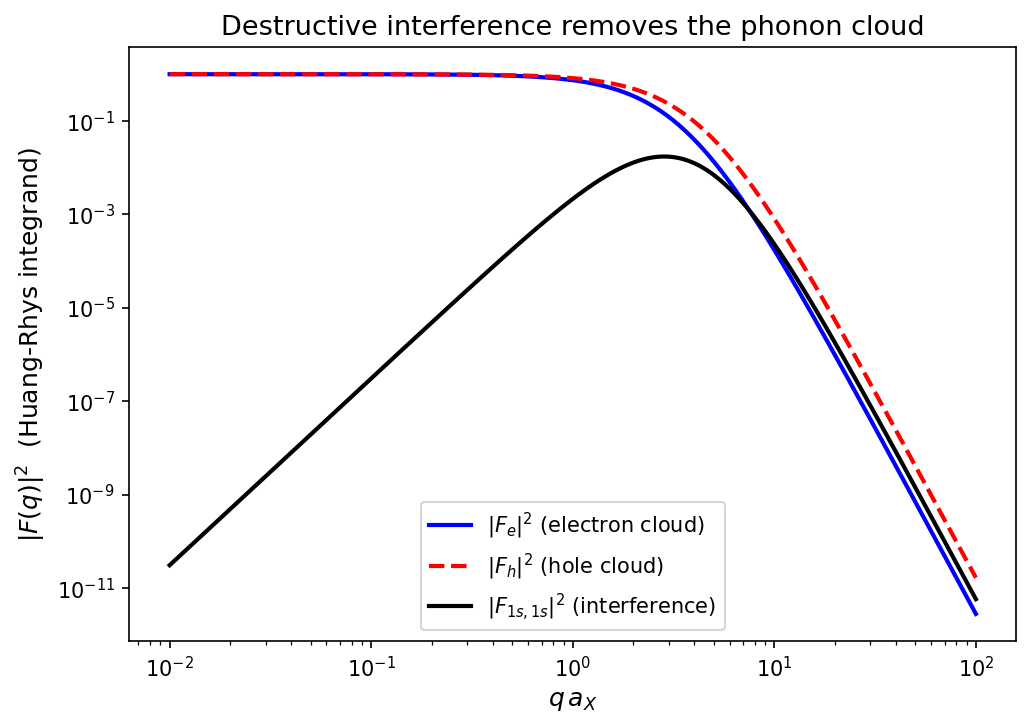}
\caption{(Color online) Huang--Rhys integrands. Destructive interference
removes the small-$q$ phonon cloud, confining the elastic dressing to
intermediate $q$.}
\label{fig:integrand}
\end{figure}

Figure~\ref{fig:formfactors} shows the form factors: the internal-state-preserving channel
(black) is suppressed by two orders of magnitude at small $q$ relative to
the constituent clouds, while the internal-state-changing channel
(magenta) retains a finite dipolar strength. Consequently the Huang--Rhys integrand
(Fig.~\ref{fig:integrand}) is confined to an intermediate-$q$ window.

\begin{figure}
\includegraphics[width=\columnwidth]{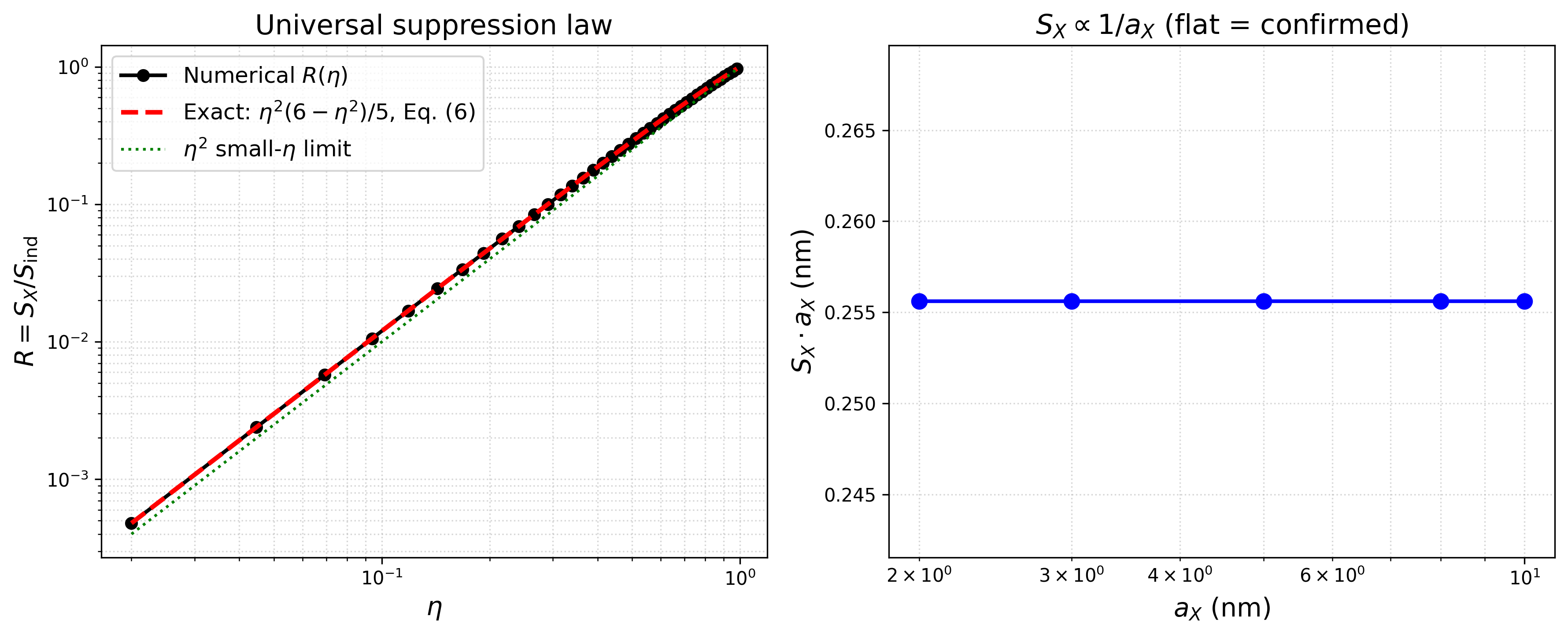}
\caption{(Color online) Universal suppression law. Left:
$R(\eta)=S_X/S_{\rm ind}$ (points) against the exact law
$\eta^2(6-\eta^2)/5$, Eq.~(\ref{eq:R}) (dashed).
Right: exact $1/a_X$ scaling of $S_X$.}
\label{fig:universal}
\end{figure}

The integral $S_X\propto\int_0^\infty dq\,|\mathcal F_{1s,1s}(q)|^2$
evaluates in closed form. With $s=\sqrt a=f_e/2$, $t=\sqrt b=f_h/2$, the
diagonal terms give $5\pi/(32s)+5\pi/(32t)$ and the cross term
\begin{equation}
\int_0^\infty\!\frac{dq}{(1+aq^2)^2(1+bq^2)^2}
=\frac{\pi\left[\sqrt{a}\,b^{3/2}(5a-b)+a^2(a-5b)\right]}{4\sqrt{a}\,(a-b)^3},
\label{eq:crossterm}
\end{equation}
whence $S_{\rm ind}\propto 5\pi(s+t)/(32st)$ and the ratio reduces exactly
to Eq.~(\ref{eq:R}). For MAPbI$_3$ parameters (Table~\ref{tab:mat}) we obtain
$S_{\rm ind}=6.7$ but $S_X=0.10$: the zero-phonon line retains
$e^{-S_X}\approx90\%$ of the spectral weight, versus $e^{-S_{\rm ind}}
\approx0.1\%$ in the naive picture. Because both integrals scale as
$1/a_X$, the ratio $R=S_X/S_{\rm ind}$ is a one-parameter function of
$\eta$ alone \emph{within the hydrogenic Fr\"ohlich model}, given exactly
by $\eta^2(6-\eta^2)/5$ [Fig.~\ref{fig:universal}].

\section{Linewidths: The Hierarchy of Channels}

\begin{figure}
\includegraphics[width=\columnwidth]{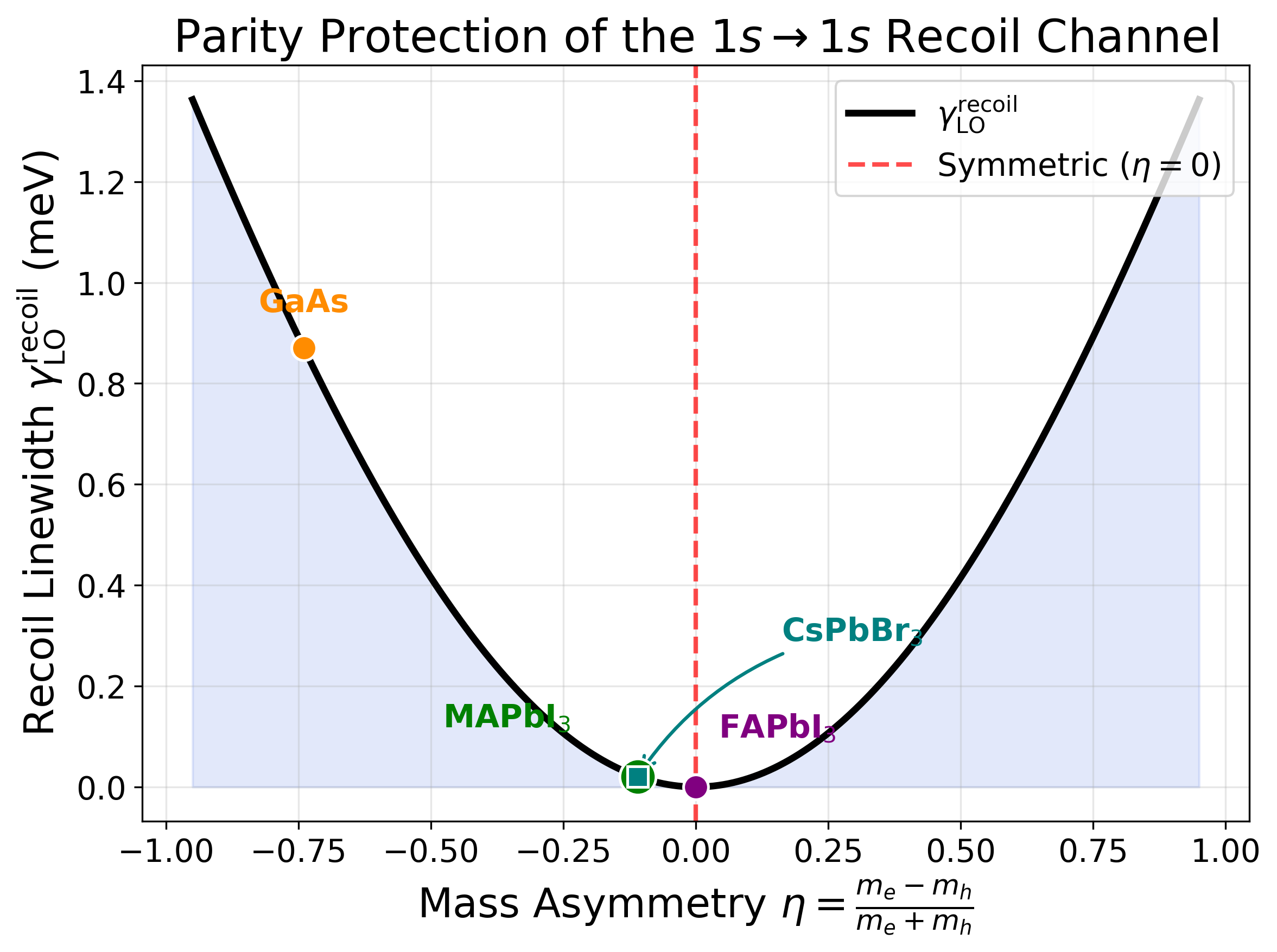}
\caption{(Color online) Parity protection of the $1s\to1s$ COM recoil
channel. The parameter-free recoil linewidth,
Eq.~(\ref{eq:recoilrate}), vanishes exactly at $\eta=0$ (mass symmetry).
Mass-symmetric perovskites sit at the protected minimum, forcing their
linewidth onto the inelastic $1s\to np$ channel; mass-asymmetric GaAs sits
on the unprotected branch.}
\label{fig:vshape}
\end{figure}

Table~\ref{tab:mat} and Fig.~\ref{fig:vshape} establish the hierarchy of exciton-phonon scattering channels governed by $(\eta, q_*a_X, \Delta E_{np}-\hbar\omega_{\rm LO})$.

\textit{Asymmetric regime (GaAs).} With $\eta=-0.74$, parity protection
fails at $q_*$ ($\mathcal F(q_*)=-0.59$), yielding the parameter-free
bound-state baseline $\gamma_{\rm LO}^{\rm recoil}=2.2$ meV. Because
$\hbar\omega_{\rm LO}>E_B$, every bound--bound absorption channel is closed
in GaAs ($\Delta E_{np}\le E_B<\hbar\omega_{\rm LO}$), so the LO-phonon
absorption linewidth is entirely ionization broadening; its definitive
treatment, with Coulomb-correlated continuum states, is
Ref.~\cite{RudinReineckeSegall1990}, and the 8--14 meV bulk values compiled
there are themselves determinations of that channel. The contribution of
the present work is the recoil baseline above, a channel absent from
frozen-COM treatments including Ref.~\cite{RudinReineckeSegall1990}. A
plane-wave (Born) evaluation of the ionization channel in the minimal model
gives $\gamma_{\rm ion}^{\rm PW}\approx110$ meV; this number is quoted only
as a formal upper bound and is not used quantitatively. Note that GaAs
($q_*a_X=8.35$) and MAPbI$_3$ (0.98) occupy dramatically different regions
of $(\eta,q_*a_X,\Delta)$: the observable is controlled by
$|\tilde M(q_*)|^2|\mathcal F(q_*)|^2\rho_{\rm COM}(q_*)$, not by $\alpha$
alone.

\textit{Symmetric regime (FAPbI$_3$).} For exact mass symmetry ($\eta=0$), Eq.~(\ref{eq:exact_zero}) dictates $\gamma_{\rm LO}^{\rm recoil} = 0$ exactly. The internal-state-preserving channel is completely closed by parity.

\textit{Near-symmetric regime (MAPbI$_3$, CsPbBr$_3$).} For MAPbI$_3$ ($\eta=-0.11$), the recoil rate is negligible ($0.12$~meV). The observed 27--40 meV linewidths therefore cannot be accounted for by
COM recoil and require internal-state-changing and/or other channels;
within the present model the $1s\to2p$ resonance, Eq.~(\ref{eq:inelastic}), is the leading
candidate. While a minimal model with a single dispersionless LO mode yields a small inelastic rate ($\sim 2$ meV), perovskites possess a broad, highly dispersive soft-phonon spectrum that massively enhances the inelastic phase space. Full \emph{ab initio} electron-phonon calculations are required to capture this absolute scale, but our model rigorously proves that the recoil channel is parity-blocked, forcing the broadening onto these inelastic channels.

The experimental column of Table~\ref{tab:mat} is organized by observable: Ref.~\onlinecite{Fu2018} fits single-exciton homogeneous linewidths; Ref.~\onlinecite{Wright2016} extracts carrier--Fr\"ohlich strengths from film PL; Ref.~\onlinecite{RudinReineckeSegall1990} provides bulk exciton--LO comparisons.

\section{Discussion}

$\alpha$ remains the natural measure of the constituent carrier--phonon
coupling, but after projection onto the correlated exciton it is necessary
rather than sufficient: the observables are fixed by
$(\eta,\ q_*a_X,\ \Delta E_{np}-\hbar\omega_{\rm LO})$. This yields
design rules: engineer $\eta\to0$ and detune the Rydberg series from
$\hbar\omega_{\rm LO}$ to protect coherence.  Limitations include the hydrogenic envelope, the
dispersionless LO mode, and the estimated continuum floor; the selection
rule itself is dimensionally invariant and extends to 2D
systems~\cite{Atambo2026a}.
Recent advances in \emph{ab initio} many-body perturbation theory have
enabled first-principles calculations of exciton--phonon matrix elements
and finite-temperature relaxation times from the Bethe--Salpeter equation
\cite{Chen2020}. In such transition-basis formulations, the exciton--phonon
matrix element is built from electron- and hole-phonon scattering amplitudes
weighted by the exciton wave function. The analytical results derived here
provide limiting cases that any such numerical calculation must satisfy:
for an exactly mass-symmetric $1s$ exciton the internal-state-preserving
Fr\"ohlich recoil matrix element vanishes identically at all $q$, whereas
the internal-state-changing $1s\to np$ dipole remains finite to leading
order. Thus the present model is not a substitute for full \emph{ab initio}
phonon dispersions and transition-basis amplitudes, but supplies exact
selection-rule benchmarks for those calculations.

\section{Conclusion}

We have shown that restoring exact center-of-mass kinematics to the exciton--phonon problem exposes a universally open, internal-state-preserving absorption channel at the recoil momentum $q_*$. The exact finite-$q$ identity $\mathcal F_{1s,1s}(q, \eta=0)=0$ proves that mass-symmetric excitons are perfectly protected from this recoil broadening. This establishes a strict hierarchy of scattering channels: monopolar COM-recoil coupling is suppressed by interference, while dipolar internal transitions and ionization channels survive. These results replace the Fr\"ohlich constant $\alpha$ by the triplet $(\eta,\ q_*a_X,\ \Delta E_{np}-\hbar\omega_{\rm LO})$ as the fundamental predictors of exciton--phonon scattering, providing rigorous design rules for coherent optoelectronic materials and establishing the necessary boundary conditions for future \emph{ab initio} electron-phonon calculations.

\begin{acknowledgments}
The author acknowledges Kenya Education Network (KENET) research services for computing resources.
\end{acknowledgments}

\begin{table*}
\caption{Regime assignment and parameter-free rates.
$q_*$: recoil momentum, Eq.~(\ref{eq:recoil});
$\mathcal{F}(q_*)$: internal-state-preserving form factor at recoil;
$\gamma_{\rm LO}^{\rm recoil}$: this work, Eq.~(\ref{eq:recoilrate});
Experimental LO-related linewidth/coupling: experimental fits, organized by observable (footnotes).}
\label{tab:mat}
\begin{ruledtabular}
\begin{tabular}{lcccccl}
 & $\eta$ & $q_*a_X$ & $\mathcal{F}(q_*)$ &
 $\gamma_{\rm LO}^{\rm recoil}$ (meV) & regime &
 $\gamma_{\rm LO}^{\rm exp}$ (meV) \\
\hline
GaAs       & $-0.74$ & 8.35 & $-0.59$ & 2.24 &
 asymmetric: recoil + ionization & 8--14$^a$ \\
MAPbI$_3$  & $-0.11$ & 0.98 & $-0.04$ & 0.12 &
 near-symmetric: inelastic resonance  & $40\pm2^b$ \\
FAPbI$_3$  & $0.00$  & 0.62 & $0.00$  & 0.00 &
 symmetric: inelastic resonance  & 27$^c$ \\
CsPbBr$_3$ & $-0.11$ & 1.32 & $-0.07$ & 0.65 &
 intermediate                    & $\sim70^d$ \\
\end{tabular}
\end{ruledtabular}
\vspace{1mm}
$^a$Bulk exciton--LO widths, Ref.~\onlinecite{RudinReineckeSegall1990};
$^b$carrier--Fr\"ohlich, Ref.~\onlinecite{Wright2016};
$^c$single-exciton linewidth, Ref.~\onlinecite{Fu2018};
$^d$PL-broadening effective coupling, Ref.~\onlinecite{CsPbBr3PL}.
\end{table*}

\begin{figure}
\includegraphics[width=\columnwidth]{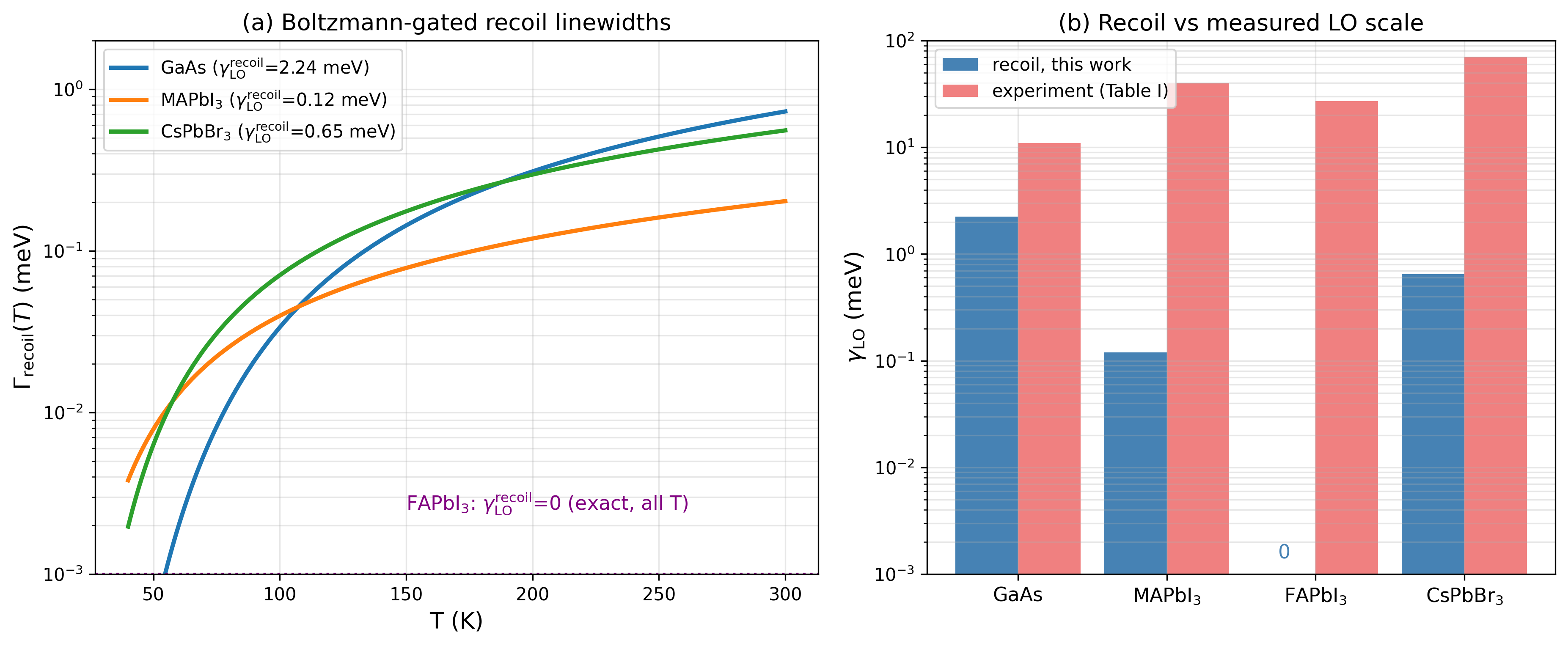}
\caption{(Color online) (a) Temperature dependence of the
internal-state-preserving recoil linewidths
$\Gamma_{\rm recoil}(T)=\gamma_{\rm LO}^{\rm recoil}N_{\rm LO}(T)$ using the
parameter-free baselines of Table~\ref{tab:mat}; for FAPbI$_3$ the rate is
exactly zero at all $T$ by Eq.~(\ref{eq:exact_zero}). (b) Recoil coefficients (blue) against
the measured LO-related scales (red, Table~\ref{tab:mat}). The recoil
channel supplies the GaAs bound-state baseline but lies two to three orders
of magnitude below the perovskite benchmarks, which therefore require
internal-state-changing channels (not shown).}
\label{fig:linewidth}
\end{figure}

\begin{thebibliography}{1}%
\makeatletter
\providecommand \@ifxundefined [1]{%
 \@ifx{#1\undefined}
}%
\providecommand \@ifnum [1]{%
 \ifnum #1\expandafter \@firstoftwo
 \else \expandafter \@secondoftwo
 \fi
}%
\providecommand \@ifx [1]{%
 \ifx #1\expandafter \@firstoftwo
 \else \expandafter \@secondoftwo
 \fi
}%
\providecommand \natexlab [1]{#1}%
\providecommand \enquote  [1]{``#1''}%
\providecommand \bibnamefont  [1]{#1}%
\providecommand \bibfnamefont [1]{#1}%
\providecommand \citenamefont [1]{#1}%
\providecommand \href@noop [0]{\@secondoftwo}%
\providecommand \href [0]{\begingroup \@sanitize@url \@href}%
\providecommand \@href[1]{\@@startlink{#1}\@@href}%
\providecommand \@@href[1]{\endgroup#1\@@endlink}%
\providecommand \@sanitize@url [0]{\catcode `\\12\catcode `\$12\catcode `\&12\catcode `\#12\catcode `\^12\catcode `\_12\catcode `\%12\relax}%
\providecommand \@@startlink[1]{}%
\providecommand \@@endlink[0]{}%
\providecommand \url  [0]{\begingroup\@sanitize@url \@url }%
\providecommand \@url [1]{\endgroup\@href {#1}{\urlprefix }}%
\providecommand \urlprefix  [0]{URL }%
\providecommand \Eprint [0]{\href }%
\providecommand \doibase [0]{https://doi.org/}%
\providecommand \selectlanguage [0]{\@gobble}%
\providecommand \bibinfo  [0]{\@secondoftwo}%
\providecommand \bibfield  [0]{\@secondoftwo}%
\providecommand \translation [1]{[#1]}%
\providecommand \BibitemOpen [0]{}%
\providecommand \bibitemStop [0]{}%
\providecommand \bibitemNoStop [0]{.\EOS\space}%
\providecommand \EOS [0]{\spacefactor3000\relax}%
\providecommand \BibitemShut  [1]{\csname bibitem#1\endcsname}%
\let\auto@bib@innerbib\@empty
\bibitem [{Note1()}]{Note1}%
  \BibitemOpen
  \bibinfo {note} {We define $\eta =(m_e-m_h)/(m_e+m_h)$, opposite in sign to the convention of Ref.~\cite {Atambo2026a}; all rates depend on $\eta ^2$ and $|F_{1s,1s}(q_*)|$ and are unaffected by this choice.}\BibitemShut {Stop}%
\end{thebibliography}%


\begin{thebibliography}{99}

\bibitem{Frohlich1937} H. Fr\"ohlich, Proc. R. Soc. London, Ser. A
\textbf{160}, 230 (1937).

\bibitem{Frohlich1954} H. Fr\"ohlich, Adv. Phys. \textbf{3}, 325 (1954).

\bibitem{HuangRhys1950} K. Huang and A. Rhys, Proc. R. Soc. London, Ser. A
\textbf{204}, 406 (1950).

\bibitem{Toyozawa1958} Y. Toyozawa, Prog. Theor. Phys. \textbf{20}, 53
(1958).

\bibitem{RudinReineckeSegall1990} S. Rudin, T. L. Reinecke, and B. Segall,
Phys. Rev. B \textbf{42}, 11218 (1990).

\bibitem{Wright2016} A. D. Wright, C. Verdi, R. L. Milot, G. E. Eperon,
M. A. P\'erez-Osorio, H. J. Snaith, F. Giustino, M. B. Johnston, and
L. M. Herz, Nat. Commun. \textbf{7}, 11755 (2016).

\bibitem{Fu2018} M. Fu, P. Tamarat, J.-B. Trebbia, M. I. Bodnarchuk,
M. V. Kovalenko, J. Even, and B. Lounis, Nat. Commun. \textbf{9}, 3318
(2018).

\bibitem{BetheSalpeter} H. A. Bethe and E. E. Salpeter, {\it Quantum
Mechanics of One- and Two-Electron Atoms} (Springer, Berlin, 1957).

\bibitem{Hedin1965} L. Hedin, Phys. Rev. \textbf{139}, A796 (1965).

\bibitem{HybertsenLouie1986} M. S. Hybertsen and S. G. Louie, Phys. Rev. B
\textbf{34}, 5390 (1986).

\bibitem{Strinati1980} G. Strinati, H. J. Mattausch, and W. Hanke,
Phys. Rev. Lett. \textbf{45}, 290 (1980).

\bibitem{Strinati1982} G. Strinati, H. J. Mattausch, and W. Hanke,
Phys. Rev. B \textbf{25}, 2867 (1982).

\bibitem{OnidaReiningRubio2002} G. Onida, L. Reining, and A. Rubio,
Rev. Mod. Phys. \textbf{74}, 601 (2002).

\bibitem{AryasetiawanGunnarsson1998} F. Aryasetiawan and O. Gunnarsson,
Rep. Prog. Phys. \textbf{61}, 237 (1998).

\bibitem{Atambo2026a} M. O. Atambo, Parity-controlled electron-hole interference in exciton-phonon coupling,  arXiv:2608.25482 (2026).

\bibitem{Atambo2026b} M. O. Atambo, Accidental accuracy and vertex corrections in $GW$: Exact benchmarks for the extended Hubbard model,  arXiv:2608.24838 (2026).

\bibitem{Atambo2026c} M. O. Atambo, Symmetry-protected intraband exciton--phonon scattering and its breakdown by mass asymmetry, arXiv:2609.29634 (2026).


\bibitem{CsPbBr3PL} C. Wolf and T.-W. Lee, Mater. Today Energy,\textbf{7},199-207 (2018).

\bibitem{Chen2020} H.-Y. Chen, D. Sangalli, and M. Bernardi, Phys. Rev. Lett. \textbf{125}, 107401 (2020).

\end{thebibliography}
\end{document}